\def\arxivversion{1}
\ifdefined\arxivversion
\documentclass[aps,prl,reprint,onecolumn,preprintnumbers,superscriptaddress,floatfix]{revtex4-2}
\else
  \documentclass[aps,prl,reprint,superscriptaddress,floatfix]{revtex4-2}
\fi
\usepackage{amsmath}
\usepackage{graphicx}
\usepackage{dcolumn}
\usepackage{bm}
\ifdefined\showrevisions
  \usepackage{xcolor}
\else
  \usepackage[monochrome]{xcolor}
\fi
\definecolor{REVISIONPURPLE}{RGB}{112,48,160}

\begin{document}

\ifdefined\arxivversion
  \preprint{RUP-26-27}
\fi

\title{Vector field dark matter search with Sagnac interferometer}

\author{Yuki Inoue}
\thanks{Corresponding author: iyuki@ncu.edu.tw}
\affiliation{Department of Physics,National Central University, Taoyuan, Taiwan}
\affiliation{Center for High Energy and High Field (CHiP), National Central University,  Taoyuan, Taiwan}
\affiliation{Institute of Physics, Academia Sinica, Taipei, Taiwan}
\affiliation{Institute of Particle and Nuclear Studies, High Energy Acceleration Research Organization (KEK), Tsukuba, Japan}
\author{Quynh Lan Nguyen}%
\affiliation{Phenikaa University, Duong Noi, Hanoi
12116, Vietnam}
\author{Ippei Obata}%
\affiliation{Institute of Particle and Nuclear Studies, High Energy Acceleration Research Organization (KEK), Tsukuba, Japan}
\affiliation{Department of Physics, Rikkyo University, Toshima, Tokyo 171-8501, Japan}
\affiliation{Kavli Institute for the Physics and Mathematics of the Universe (Kavli IPMU), WPI, UTIAS, The University of Tokyo, Kashiwa, Chiba 277-8568, Japan}

\date{\today}

\begin{abstract}
We present the first calculation showing that a material-asymmetric Sagnac speed meter can improve quantum-noise-limited sensitivity to $B-L$ vector dark matter by nearly two orders of magnitude over a Fabry--Perot Michelson position meter at the low-mass end of the band. We introduce material contrast into a dual-ring interferometer to probe composition-dependent $B-L$ forces and compare the two configurations at equal incident power, input-mirror transmissivity, and cavity round-trip length. For sapphire mirrors and a one-week benchmark, the quantum-noise-only median expected $95\%$ CL upper limit reaches $1.2\times10^{-27}$ at $m_A=4.1\times10^{-14}\,\mathrm{eV}/c^2$ ($10\,\mathrm{Hz}$), a factor-$57$ improvement. This band-edge forecast assumes negligible classical noise; sensitivity at $10\,\mathrm{Hz}$ is not established for a realizable suspension and isolation system. The material-induced displacement is essentially identical in both configurations, with the gain arising from reduced low-frequency quantum backaction. These results identify speed-meter readout as a promising route to lower-mass vector-dark-matter searches.
\end{abstract}

\maketitle


\paragraph{Introduction.---}

The nature of dark matter (DM) remains a central open question in modern physics. Among the DM candidates, ultralight DM is particularly intriguing because its large de Broglie wavelength allows it to behave as a coherently oscillating classical field on laboratory scales \cite{HuiEtAl2017,Ferreira:2020fam}. Such a field can exert an oscillating force on ordinary matter, enabling direct laboratory searches.

Vector DM can induce a force set by the field polarization, leading to oscillatory displacements of macroscopic test masses. A well-motivated example is a $U(1)_{B-L}$ gauge boson, which couples to the baryon minus lepton ($B-L$) charge of ordinary matter and produces a composition-dependent force \cite{GrahamAccelerometers2016,Manley:2020mjq}. Laser interferometric gravitational-wave detectors provide a sensitive platform for such searches through oscillations of test-mass separations \cite{PierceRilesZhao2018,GuoEtAl2019,Michimura:2020vxn,MorisakiEtAl2021,Michimura:2025kod}. Searches using Advanced LIGO~\cite{Aasi2015AdvancedLIGO}, Advanced Virgo~\cite{Acernese2015AdvancedVirgo}, and KAGRA~\cite{KAGRA2021Overview} have already constrained ultralight vector DM over a broad mass range corresponding roughly to the audio band of ground-based interferometers \cite{LIGOScientific:2021ffg,AbbottEtAl2024DarkPhoton}. At lower frequencies, however, conventional interferometers are increasingly limited by technical and environmental noise, including seismic and Newtonian noise, while quantum radiation-pressure noise also becomes important. This motivates interferometric techniques with improved low-frequency sensitivity.

A speed meter is a promising quantum-nondemolition scheme that suppresses quantum back-action by measuring the velocity rather than the instantaneous position of a test mass. The Sagnac interferometer naturally realizes such a speed meter and can significantly reduce quantum radiation-pressure noise at low frequencies \cite{Chen2003Sagnac,Graf2014Speedmeter,PurdueChen2002Speedmeter,InoueEtAl2026CHRONOS}. Here we investigate the sensitivity of a Sagnac speed-meter interferometer to ultralight vector DM, focusing on the $U(1)_{B-L}$ gauge boson. We demonstrate improved low-frequency sensitivity and explore the resulting constraints on the $B-L$ gauge coupling.

\begin{figure}[t]
	\centering
	\includegraphics[width=\columnwidth]{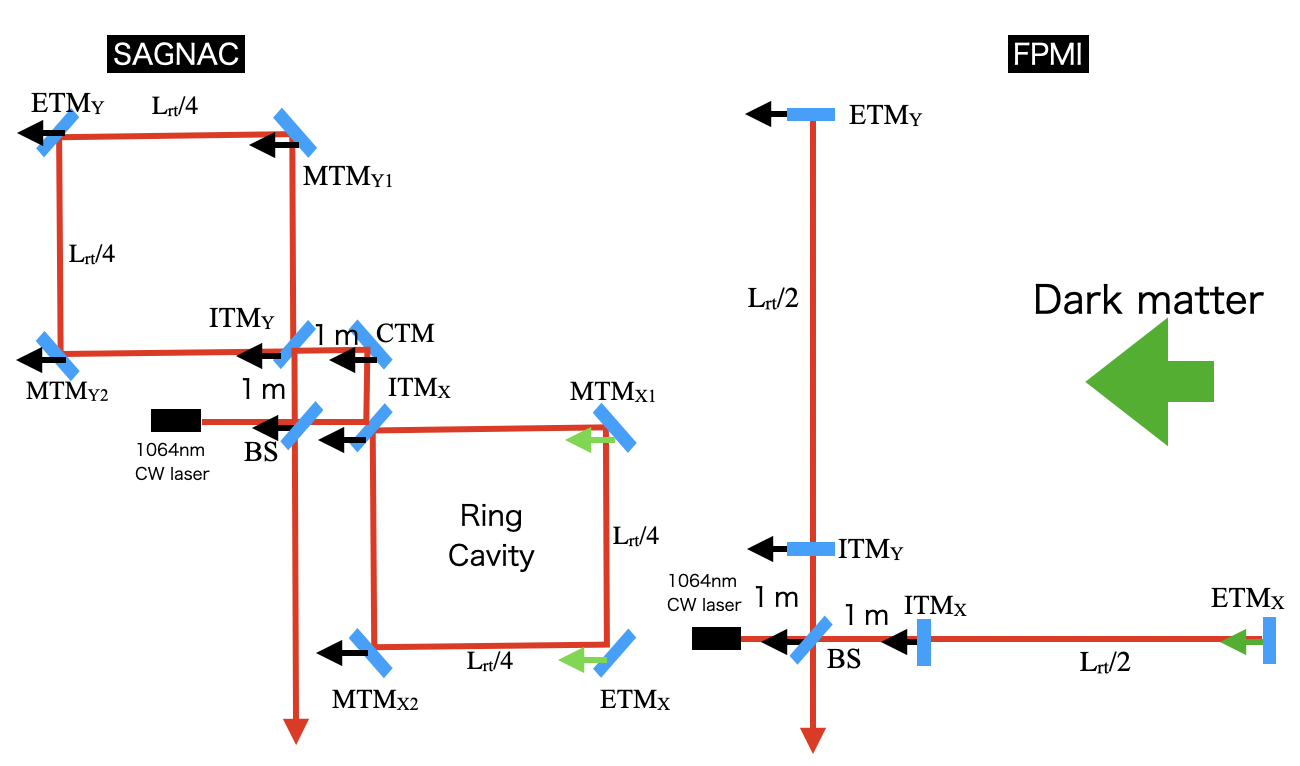}
	\caption{Optical configurations of the Sagnac speed
		meter (left) and FPMI position meter (right).
		The two configurations are compared at the same
		cavity round-trip length,
		$L_{\rm rt}=10\,\mathrm{m}$:
		each square-ring side is
		$L_{\rm rt}/4=2.5\,\mathrm{m}$,
		whereas each FP arm has a one-way length of
		$L_{\rm rt}/2=5\,\mathrm{m}$.
		The BS--ITM links and the ITM--CTM links are
		$1\,\mathrm{m}$ each.
		$\mathrm{MTM}_{X1}$ and $\mathrm{ETM}_{X}$ in Sagnac, and $\mathrm{ETM}_{X}$ in FPMI,
		are sapphire or silicon; the other cavity mirrors
		are fused silica.
		Routing optics and the homodyne reference
		are fixed in the model.}
	\label{fig:experimental_configuration}
\end{figure}

\paragraph{Model.---}
In this section, we present a schematic setup of Sagnac speed meter to probe the oscillatory fources induced by the $B-L$ dark matter.
Figure~\ref{fig:experimental_configuration} shows the dual-ring topology of Sagnac interferometer. 
The mirror labels denote
End Test Mass ($\mathrm{ETM}_{l}$),
Input Test Mass ($\mathrm{ITM}_{l}$),
Middle Test Mass ($\mathrm{MTM}_{l}$),
and Center Test Mass ($\mathrm{CTM}_{l}$),
where $l=\{X,Y,X1,X2,Y1,Y2\}$ identifies the mirror;
BS denotes the Beam Splitter.
The single center mirror
is labeled CTM without a subscript
in Fig.~\ref{fig:experimental_configuration}.
$\mathrm{MTM}_{X1}$ and $\mathrm{ETM}_{X}$ are both sapphire or silicon; all other cavity mirrors are fused silica (FS). In FPMI only $\mathrm{ETM}_{X}$ is replaced. All cavity mirrors remain $40$-kg free masses, and material substitution changes only the charge-to-mass parameter. The laser reference, and ideal external local oscillator (LO) are fixed. Thus this model does not describe an apparatus whose entire reference system responds freely to the dark-matter field.

In natural units ($\hbar=c=1$), the interaction and peak acceleration are
\begin{align}
\mathcal L_{\rm int}&=-\epsilon_{B-L}e A_\mu J_{B-L}^\mu,\nonumber\\
\bm a_{j,0}&=\epsilon_{B-L}e\frac{\eta_j}{m_n}\sqrt{2\rho_{\rm DM}}\,\bm e_A,
\label{eq:interaction_lagrangian}
\end{align}
Here $\mathcal L_{\rm int}$ is the interaction Lagrangian density, $A_\mu$ the vector-DM field, and $J_{B-L}^\mu$ the baryon-minus-lepton four-current, with the repeated index $\mu$ contracted. The dimensionless coupling $\epsilon_{B-L}$ is given in units of $e=\sqrt{4\pi\alpha_{\rm EM}}$, where $\alpha_{\rm EM}$ is the fine-structure constant. For mirror $j$, $M_j$ and $Q_j$ denote its mass and dimensionless total $B-L$ charge, and $\eta_j=m_nQ_j/M_j$ is the material charge-to-mass parameter, with $m_n$ the neutron mass. For electrically neutral matter, proton and electron contributions cancel, so $Q_j$ equals the neutron number. Here $\rho_{\rm DM}$ is the local vector-DM energy density, $\bm a_{j,0}$ the peak acceleration, and $\bm e_A$ the unit polarization vector. The factor $\sqrt{2}$ converts the time-RMS field amplitude to its peak value. We assume a spatially uniform field with an isotropic three-dimensional polarization distribution. In SI units,
\begin{align}
\bm a_{j,0}&=\epsilon_{B-L}\eta_j a_*\bm e_A,\nonumber\\
a_*&=\frac{\sqrt{8\pi\alpha_{\rm EM}\hbar c\rho_{\rm DM}^{\rm SI}}}{m_n}
={3.6\times10^{11}}\,\mathrm{m\,s^{-2}},
\label{eq:si_acceleration}
\end{align}
for the fiducial dark-matter density in Table~\ref{tab:parameters}. The bare acceleration-to-displacement susceptibility is $\chi_a=-\Omega^{-2}$, where $\Omega=2\pi f$ is the angular frequency of the force. For illustration, a single undamped pendulum with $f_0=0.60\,\mathrm{Hz}$ has $\chi_{a,p}=[(2\pi f_0)^2-\Omega^2]^{-1}$, differing in magnitude from the free-mass response by $(f_0/f)^2/[1-(f_0/f)^2]=0.36\%$ at $10\,\mathrm{Hz}$. This estimate concerns only the response to a force on the test mass; it neither bounds support-motion noise nor models a multistage suspension. The $10$--$5000\,\mathrm{Hz}$ band is a benchmark for the ideal free-mass comparison, not a demonstrated low-noise operating band.

Let $L_{\rm rt}$ denote the geometric round-trip length
of the ring cavity, and let $\bm n_j$ be a unit vector
normal to the surface of mirror $j$, with its orientation
chosen to define the positive local displacement.
A small normal displacement of this mirror is
$\delta\bm r_j=z_j\bm n_j$, where $z_j$ has units of length. $\bm u_{{\rm in},j}$ and $\bm u_{{\rm out},j}$
denote unit vectors along the incident and reflected
light propagation directions, respectively.
The displacement changes the lengths of the two
adjacent optical segments, giving the first-order
round-trip length perturbation
\begin{equation}
\delta L_{\rm rt}
=\sum_j
(\bm u_{{\rm in},j}-\bm u_{{\rm out},j})
\cdot\bm n_j\,z_j.
\end{equation}
Here $\delta L_{\rm rt}$ is the first-order change in the geometric round-trip path length of one cavity relative to its unperturbed value $L_{\rm rt}$. The symbol $\delta$ denotes a small perturbation, and the sum runs over all mirrors of that cavity.
For specular reflection, this becomes
\begin{equation}
\delta L_{\rm rt}
=\sum_j 2s_j\cos i_j\,z_j,
\end{equation}
where $i_j$ is the incidence angle measured from the
surface normal and $s_j=\pm1$ accounts for the chosen
orientation of $\bm n_j$.
Thus a square ring has a displacement-to-length
factor of magnitude $\sqrt{2}$ at each mirror
($i_j=45^\circ$), whereas a normally incident
Fabry--Perot cavity has a factor of magnitude $2$.
For compact notation, we define
\begin{equation}
\bm g_j=\bm u_{{\rm in},j}-\bm u_{{\rm out},j}.
\end{equation}
This dimensionless vector is a geometric response
coefficient, not a unit normal vector.
The expression above describes the instantaneous
geometric length change; finite propagation and
cavity-storage effects are included separately
in the frequency-dependent optical response.
We calculate that response using Finesse~3~\cite{Finesse3},
an interferometer simulation package.
The signed reflection factors are included in its
optical response and are not applied again to the
external force, which is projected onto $\bm n_j$.

We use the common displacement-calibration coordinate
\begin{equation}
q=\frac{\delta L_{{\rm rt},X}
       -\delta L_{{\rm rt},Y}}{2},
\label{eq:differential_ring_length}
\end{equation}
which reduces to the differential one-way arm-length
change in FPMI. This is a geometric calibration
coordinate; the measured output additionally depends
on the frequency-dependent optical response.
For local normal displacements $z_j$, we have
$q=\sum_j c_jz_j$, with
\begin{equation}
c_j=\frac{\sigma_j}{2}\bm g_j\cdot\bm n_j
   =\sigma_j s_j\cos i_j,
\end{equation}
where $\sigma_j=+1$ for the X cavity and $-1$ for
the Y cavity. All sums run over the mirrors in
both arm cavities.
To define the calibration response, we impose the
displacement pattern
\begin{equation}
z_j=\frac{c_j}{\sum_k c_k^2}q,
\qquad
H_q=\sum_j H_{yz_j}\frac{c_j}{\sum_k c_k^2},
\label{eq:calibration_pattern}
\end{equation}
where $H_{yz_j}$ is the complex output response
per imposed local displacement.
This calibration pattern need not coincide with
the displacement pattern driven by dark matter.
For equal mirror masses $M$, the effective mass
associated with this coordinate is
$M_{\rm eff}=M/\sum_j c_j^2$.
The square-ring Sagnac has eight mirrors with
$c_j^2=1/2$, whereas FPMI has four mirrors with
$c_j^2=1$. Both therefore give
$M_{\rm eff}=M/4=10\,\mathrm{kg}$.

Choose the in-plane unit vector $\bm e_x$ such that
the two substituted Sagnac mirrors satisfy
$\bm g_{\rm MTMX1}+\bm g_{\rm ETMX}=2\bm e_x$,
and define the angle $\phi$ between the field polarization $\bm e_A$
and the material-contrast axis $\bm e_x$ by
$\bm e_A\cdot\bm e_x=\cos\phi$.
With $\Delta\eta=\eta_{\rm mat}-\eta_{\rm FS}$,
the bare material-contrast response is
\begin{equation}
\frac{q_{\rm contrast}}{\epsilon_{B-L}}
=a_*\Delta\eta\,\chi_a\cos\phi.
\end{equation}
The quantity $q_{\rm contrast}$ is the material-contrast contribution to the differential displacement $q$: the bare geometric response with the selected mirrors replaced, minus that for an otherwise identical all-fused-silica configuration. Thus $q_{\rm contrast}/\epsilon_{B-L}$ is displacement per unit coupling. Here $a_*$ is the acceleration scale per unit $\epsilon_{B-L}\eta_j$ defined in Eq.~(\ref{eq:si_acceleration}), and the free-mass susceptibility $\chi_a$ converts acceleration into displacement. This expression precedes the optical propagation and cavity-storage response; it is not the full measured DM signal.
This equals the corresponding response of the
substituted FPMI end mirror when its displacement
axis is aligned with $\bm e_x$.
The equality concerns the bare geometric
material-contrast response, not the complete
optical transfer functions.

\begin{table}[t]
\caption{Principal parameters shared by Sagnac and FPMI.
The round-trip length corresponds to a $2.5$-m
square-ring side or a $5$-m FP arm.}
\label{tab:parameters}
\begin{ruledtabular}
\begin{tabular}{lll}
Parameter & Symbol & Value \\
\hline
Laser wavelength
& $\lambda$ & $1064\,\mathrm{nm}$ \\
Incident optical power
& $P_{\rm in}$ & $180\,\mathrm{W}$ \\
ITM power transmissivity
& $T_{\rm ITM}$ & $500\,\mathrm{ppm}$ \\
Cavity round-trip length
& $L_{\rm rt}$ & $10\,\mathrm{m}$ \\
Cavity finesse
& $\mathcal F$ & $1.3\times10^4$ \\
Individual mirror mass
& $M_j=M$ & $40\,\mathrm{kg}$ \\
Local dark-matter density
& $\rho_{\rm DM}$ & ${\color{blue}0.40}\,\mathrm{GeV\,cm^{-3}}$ \\
Halo velocities
& $v_{\rm vir},v_\odot$ & $220,232\,\mathrm{km\,s^{-1}}$ 
\end{tabular}
\end{ruledtabular}
\end{table}

\paragraph{Optical response and statistics.---}
We calculate the force response and one-sided quantum-noise spectrum with Finesse~3 for the parameters in Table~\ref{tab:parameters}. Both interferometers are resonant and lossless, with a 50:50 main beam splitter and perfectly reflecting non-input cavity mirrors. A coherent carrier and unsqueezed vacuum fluctuations are assumed. Ideal balanced homodyne detection uses the signal phase quadrature and a $1\,\mathrm{W}$ external LO. The accompanying Finesse~3.0.2 code specifies the optical connections and evaluates the complex transfer functions and quantum noise. The force on each cavity mirror and the resulting output response are
\begin{equation}
\frac{F_j}{\epsilon_{B-L}}=M_j a_*\eta_j\bm n_j\cdot\bm e_A,
\qquad H_\epsilon=\sum_jH_{yF_j}\frac{F_j}{\epsilon_{B-L}}.
\label{eq:full_response}
\end{equation}
The transfer functions include propagation delays, cavity storage, reflection geometry, and free-mass radiation-pressure feedback. No additional analytic Sagnac factor is applied. The cavity pole is $cT_{\rm ITM}/(4\pi L_{\rm rt})\simeq1.2\,\mathrm{kHz}$; thus the small single-pass propagation phase does not justify neglecting the storage response over this band.

We use the polarization RMS response
\begin{equation}
H_{\epsilon,\mathrm{rms}}
=\sqrt{\langle|H_\epsilon|^2\rangle}
=\sqrt{\frac{|H_{\epsilon,x}|^2+|H_{\epsilon,y}|^2}{3}},
\label{eq:polarization_rms}
\end{equation}
where $H_{\epsilon,a}$ is Eq.~(\ref{eq:full_response}) for polarization along axis $a$, and $H_{\epsilon,z}=0$ in this planar model.

The output quantum-noise ASD $S_y^{1/2}$ includes shot noise, radiation-pressure backaction, and their correlations. The equivalent coupling and differential-displacement noises are
\begin{equation}
S_\epsilon^{1/2}=S_y^{1/2}/H_{\epsilon,\mathrm{rms}},
\qquad S_q^{1/2}=S_y^{1/2}/|H_q|,
\end{equation}
where $H_q$ is the optical response to the common differential coordinate. Thus $S_\epsilon^{1/2}$ has units $\mathrm{Hz}^{-1/2}$, whereas $S_q^{1/2}$ has units $\mathrm{m\,Hz}^{-1/2}$. We express the median expected upper limit as $\epsilon_{95}=S_\epsilon^{1/2}\mathcal I_{95}$, where $\mathcal I_{95}$ has units $\mathrm{s}^{-1/2}$.
The distinction between the two readouts is clearest below the cavity bandwidth. In the free-mass regime a sinusoidal force produces a displacement proportional to $f^{-2}$. A position meter measures this displacement, whereas the low-frequency Sagnac output measures a difference between positions sampled at different times, and therefore carries an additional factor proportional to $f$. Its force-to-output response consequently scales approximately as $f^{-1}$. These signal scalings should not be confused with the displacement-equivalent shot-noise spectrum: a weaker low-frequency displacement response increases the input-equivalent shot noise. The speed-meter advantage instead arises from the changed radiation-pressure backaction and its correlations with the measured light. The total quantum-noise spectrum, rather than the signal transfer function alone, determines the coupling sensitivity.

The optical power is not adjusted to make the shot-noise curves coincide. Equal input power yields different total stored energies: $0.096$ J for Sagnac and $0.048$ J for FPMI. Each traveling-wave circulating power is about $7.2\times10^2\,\mathrm{kW}$. Equal incident power thus differs from equal stored energy; the latter comparison would require recalculating both noise spectra.

We adopt the KAGRA O3GK halo and field assumptions~\cite{KAGRA2024VectorDM}: the fiducial density in Table~\ref{tab:parameters}, an untruncated Maxwell distribution with $v_{\rm vir}=220\,\mathrm{km\,s^{-1}}$ and $v_\odot=232\,\mathrm{km\,s^{-1}}$, and Gaussian isotropically polarized vector components independent of velocity. For a continuous seven-day observation with stationary quantum noise, we construct a pointwise median expected $95\%$ upper limit from the finite-observation signal covariance, without imposing KAGRA's segment length or excess-power bin width.

In the noise-whitened covariance eigenbasis, let $\lambda_k$ satisfy $\sum_k\lambda_k=1$ and define $Q=\sum_k\lambda_k|z_k|^2$, with $\langle|z_k|^2\rangle=1+a\lambda_k$ and $a=\epsilon^2H_{\epsilon,\mathrm{rms}}^2T_{\rm obs}/(2S_y)$. We define
\begin{equation}
F_Q(Q_{0,\mathrm{med}};a_{95})=0.05,\qquad
\mathcal I_{95}=\sqrt{2a_{95}/T_{\rm obs}},
\end{equation}
where $Q_{0,\mathrm{med}}$ is the noise-only median. The CDF is evaluated for a weighted sum of independent exponential variables. We use finite-time covariance for $T_{\rm obs}f(v_{\rm vir}/c)^2\leq64$ and a stationary spectral approximation above it; the transition changes the forecast by less than $0.1\%$.

Because the installation site is unspecified, we average the rotating antenna covariance over detector orientations, multiplying the temporal kernel by $[1+2\cos(2\pi\Delta t/T_{\rm sid})]/3$, with $T_{\rm sid}=86164.09\,\mathrm{s}$. This is a Gaussian forecast using the averaged covariance, not an average of site-specific limits. A real search requires the actual antenna response, duty cycle, and noise; published KAGRA limits are retained without rescaling.

Seven days is a common comparison benchmark, not an optimized observation time or special DM timescale. With the KAGRA convention, $\tau_c=[f(v_{\rm vir}/c)^2]^{-1}$ is about $1.9\times10^5\,\mathrm{s}$ at $10\,\mathrm{Hz}$ and $370\,\mathrm{s}$ at $5\,\mathrm{kHz}$. The ideal amplitude scalings are $T_{\rm obs}^{-1/2}$ for $T_{\rm obs}\ll\tau_c$ and $(T_{\rm obs}\tau_c)^{-1/4}$ for $T_{\rm obs}\gg\tau_c$~\cite{GrahamAccelerometers2016,Nakatsuka:2022gaf}. Changing the common observation time alters the absolute reach but not the Sagnac-to-FPMI sensitivity ratio.

\begin{figure}[t]
\centering
\includegraphics[width=\columnwidth]{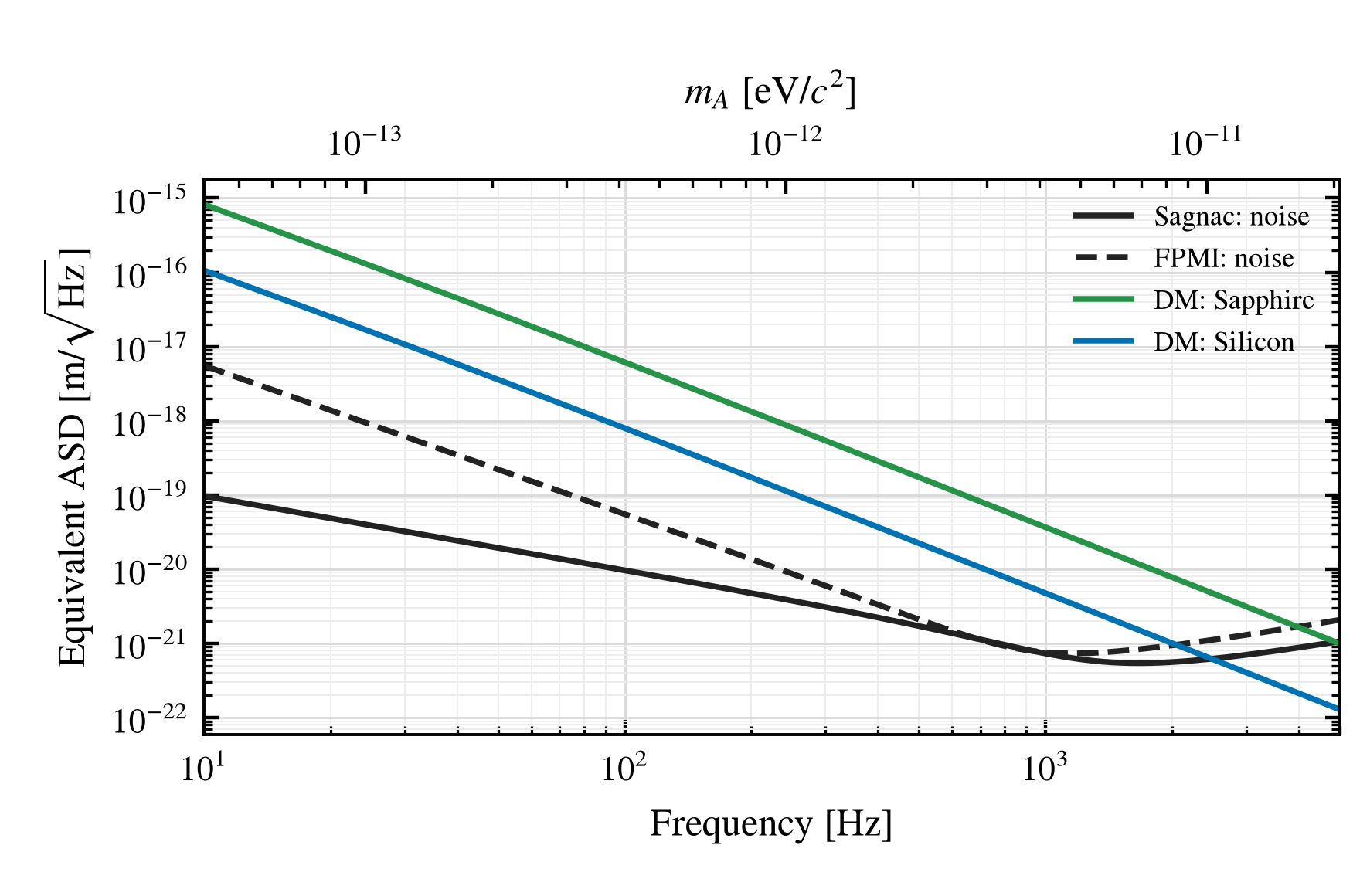}
\caption{Displacement-equivalent quantum noise of Sagnac (black solid) and FPMI (black dashed). Green and blue show sapphire and silicon signals, $\epsilon_{\rm ref}H_{\epsilon,\mathrm{rms}}/(|H_q|\mathcal I_{95})$, for $\epsilon_{\rm ref}=10^{-23}$, the fiducial density in Table~\ref{tab:parameters}, and seven days. The 3D polarization is isotropic and the rotating antenna covariance is orientation-averaged. Each material signal is shown once because the two optical configurations agree within $5$ ppm. The effective ordinate is not a physical DM ASD.}
\label{fig:strain_noise_and_dm_signal}
\end{figure}

\paragraph{Results.---}
Figure~\ref{fig:strain_noise_and_dm_signal} compares the noise with $A_{q,\rm eff}=\epsilon_{\rm ref}H_{\epsilon,\mathrm{rms}}/(|H_q|\mathcal I_{95})$, whose ratio to $S_q^{1/2}$ is $\epsilon_{\rm ref}/\epsilon_{95}$. The reduced advantage near the cavity bandwidth and the separation at higher frequencies reflect the different optical transfer functions. 
The mass axis gives $m_A=hf/c^2$.

The apparent knee of the Sagnac noise around several hundred hertz is smooth, not a discontinuity or a suspension resonance. The displacement-calibrated spectrum combines the frequency dependence of the optical response with radiation-pressure backaction and shot noise as the cavity bandwidth is approached. In the quantum-noise decomposition, the frozen-mirror shot-noise baseline contributes about $16\%$ of the total noise power at $500\,\mathrm{Hz}$; it becomes comparable to the remaining optomechanical contribution near $1.1\,\mathrm{kHz}$. Thus $500\,\mathrm{Hz}$ is not the shot/backaction equality frequency. A denser evaluation across this region confirms a smooth change of slope. Neither the one-week integration factor nor a change in the DM coherence regime enters the plotted noise ASD.

The material parameters are $\eta_{\rm FS}\simeq0.50560$, $\eta_{\rm sapphire}\simeq0.51455$, and $\eta_{\rm silicon}\simeq0.50670$ (all dimensionless) for representative natural-isotope compositions. Five significant figures retain the small silicon--FS contrast without implying a measured sample composition; the calculation uses unrounded values. The dimensionless contrasts are $\Delta\eta_{\rm sapphire}\simeq8.9\times10^{-3}$ and $\Delta\eta_{\rm silicon}\simeq1.1\times10^{-3}$. The contrast ratio $\Delta\eta_{\rm sapphire}/\Delta\eta_{\rm silicon}\simeq8.1$ applies to the subtracted response $H_{\epsilon,\rm mat}-H_{\epsilon,\rm all\ FS}$. The figures instead use the orientation RMS of the full response, giving a silicon-to-sapphire sensitivity ratio of $7.7$. A common-material residual relative to the fixed reference optics remains at finite frequency, despite cancellation of the instantaneous geometric perimeter term.

\begin{figure}[t]
\centering
\includegraphics[width=\columnwidth]{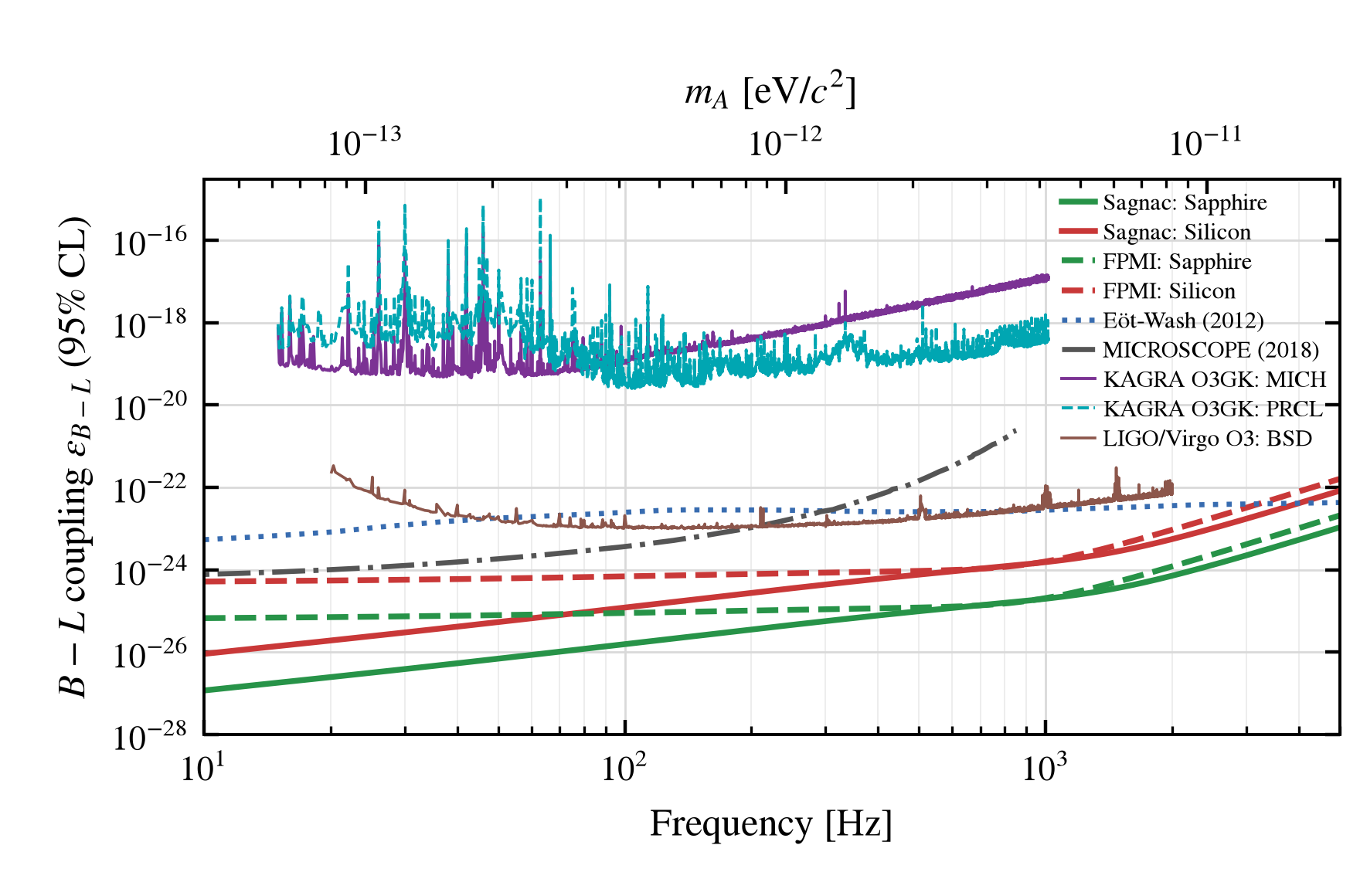}
\caption{One-week median expected $95\%$ upper limits for sapphire (green) and silicon (red), with Sagnac solid and FPMI dashed, using the fiducial density in Table~\ref{tab:parameters} and the continuous Gaussian-field forecast described in the text. Purple solid and cyan dashed curves are the published observed KAGRA O3GK MICH and PRCL limits~\cite{KAGRA2024VectorDM}, without rescaling or extrapolation. Blue dotted and gray dash-dotted curves are historical E\"ot-Wash (2012) and MICROSCOPE (2018) finite-range force benchmarks from Ref.~\cite{BergeEtAl2018}; these do not assume a local DM density. Forecasts include quantum noise only.}
\label{fig:vector_dm_sensitivity}
\end{figure}

Figure~\ref{fig:vector_dm_sensitivity} also shows
existing constraints on the $B-L$ coupling from
E\"ot-Wash and MICROSCOPE, as presented in
Ref.~\cite{BergeEtAl2018}. 
The same coupling convention is used as
Eq.~(\ref{eq:interaction_lagrangian}) in these articles.
These constraints arise from tests of
composition-dependent forces and do not require
the vector field to constitute the local dark matter.
Our projected sensitivity, in contrast, assumes
the fiducial dark-matter density in Table~\ref{tab:parameters}
and a one-week observation.
The comparison indicates the coupling range
accessible under these assumptions.
The Sagnac and FPMI curves are median expected $95\%$ CL upper limits under the assumed noise and signal model, whereas the KAGRA curves are limits from observed data.
One can see that the Sagnac interferometer has the potential to improve the sensitivity for vector boson masses below ($m_A \lesssim 10^{-11}$) eV. This mass range also partially overlaps with the region constrained by astrophysical observations, including black hole superradiance \cite{Baryakhtar:2017ngi,East:2017mrj,Cardoso:2018tly,LIGOScientific:2025csr,LIGOScientific:2026pwx}. We therefore expect that combining ground-based laboratory experiments with astrophysical probes will enable a comprehensive exploration of this mass range.

Other relevant approaches include composition-sensitive torsion balances and atom interferometers~\cite{GrahamAccelerometers2016}, and resonant mechanical force sensors~\cite{CarneyEtAl2019}. These probe oscillating differential forces, with different bandwidth, integration time, and environmental-noise requirements. A cavity-optomechanical search has now demonstrated vector-DM constraints near $39\,\mathrm{kHz}$~\cite{Chowdhury2026}, outside the present $10$--$5000\,\mathrm{Hz}$ band; its limit is therefore not extrapolated onto Fig.~\ref{fig:vector_dm_sensitivity}. KAGRA's auxiliary-length search~\cite{KAGRA2024VectorDM} is a particularly direct experimental comparison because it also exploits mirror-composition contrast. The plotted MICROSCOPE (2018) curve is a historical benchmark, not the final mission result~\cite{Touboul2022Final}; updating its finite-range $B-L$ interpretation requires the appropriate source and charge model. Static equivalence-principle bounds and DM searches have distinct density assumptions, so a quantitative comparison must match coupling conventions, mass range, halo assumptions, and statistical definitions.

Classical noise may be significant at $10\,\mathrm{Hz}$. Seismic and suspension thermal noise, as well as environmental gravity fluctuations not removed by mechanical isolation, can limit the measurement~\cite{Coughlin2016Newtonian}. Control noise and other suspension modes also require apparatus-specific assessment. The single-pendulum estimate above does not constrain these effects. For independent classical noise, $S_{q,\mathrm{tot}}=S_{q,\mathrm{quant}}+S_{q,\mathrm{class}}$, so the projected coupling limit degrades by $\sqrt{1+S_{q,\mathrm{class}}/S_{q,\mathrm{quant}}}$. Thus the factor-$57$ advantage at $10\,\mathrm{Hz}$ is quantum-noise-only; experimental realization requires suspension transfer functions and a complete noise budget.

The isotropic field and orientation-averaged antenna covariance are ensemble assumptions. A specified installation must use its actual rotating antenna vectors in the covariance.

The uniform-field approximation can be checked independently of the optical storage time. For a characteristic speed $v_0\sim10^{-3}c$, the spatial phase across a baseline $L$ is $kL\simeq2\pi f(v_0/c)L/c$. Even taking $L=10\,\mathrm{m}$ at $f=5000\,\mathrm{Hz}$ gives $kL\simeq1.0\times10^{-6}$. Spatial gradients are therefore small for the material-contrast channel considered here. This does not imply a negligible optical response time: the repeated photon passages accumulate a storage delay, which remains in the transfer functions. The two approximations concern different physical scales.

\paragraph{Conclusion.---}
In this letter, we have shown for the first time that, for a
configuration with a quantum-noise knee around
$1\,\mathrm{kHz}$, a material-asymmetric Sagnac
speed meter improves the coupling sensitivity
to vector dark matter in the low-mass regime
relative to FPMI at equal input power and
ITM transmissivity. For the one-week benchmark, the quantum-noise-only median expected $95\%$ CL upper limit reaches $1.2\times10^{-27}$ at $10\,\mathrm{Hz}$; this is conditional on classical noise being subdominant. The improvement reaches
a factor of $57$, nearly two orders of magnitude, corresponding to
$m_A=4.1\times10^{-14}\,\mathrm{eV}/c^2$,
the lower boundary of the chosen analysis band, rather than a theoretically preferred mass.
An experimental implementation will require
an assessment of environmental noise and
mirror-coating quality, including optical
losses and coating thermal noise.
These issues lie beyond the scope of the
present quantum-noise-limited comparison
and are left for future work.
\begin{acknowledgments}
\paragraph{Acknowledgments.---}
We thank Masashi Hazumi for helpful comments and suggestions that improved the clarity of this work.
Y.I. acknowledges support from NSTC, and CHiP in Taiwan under Grant No.115-2112-M-008 -022 -; Q.L.N acknowledges support NAFOSTED Grant No.103.01-2025.147; I.O. {\color{black}acknowledges} support from JSPS KAKENHI Grant No. 26K22339 and by KEK support program for young {\color{black}researchers} No. B0E510521.
\end{acknowledgments}

\bibliography{apssamp}

\end{document}